\pdfoutput=1
\documentclass{conm-p-l}

\usepackage{amssymb}
\usepackage{graphicx}
\usepackage{bm}

\copyrightinfo{2014}{American Mathematical Society}

\theoremstyle{definition}

\theoremstyle{remark}

\newcommand{\IGNORE}[1]{}
\newcommand{\ignore}[1]{}
\newcommand{\veps}{\varepsilon}

\newcommand{\opn}{\operatorname}
\newcommand{\re}{\operatorname{Re}}
\newcommand{\im}{\operatorname{Im}}

\newcommand{\mc}[1]{\mathcal{#1}}

\newcommand{\jt}{\textstyle}

\newcommand{\der}[2]{\frac{\partial #1}{\partial #2}}

\newcommand{\wdg}[1]{{#1}^{\scriptscriptstyle\bm\wedge}}

\numberwithin{equation}{section}

\begin{document}

\title[Elastic collisions of water waves]{
Relative-Periodic Elastic Collisions of Water Waves}

\author[J. Wilkening]{Jon Wilkening}
\address{Dept of Mathematics, University of California, Berkeley, CA 94720-3840}
\email{wilkening@berkeley.edu}
\thanks{This research was supported in part by the Director, Office of
Science, Computational and Technology Research, U.S.  Department of
Energy under Contract No.~DE-AC02-05CH11231, and by the National
Science Foundation through grant DMS-0955078.}


\date{April 21, 2014}

\begin{abstract}
  We compute time-periodic and relative-periodic solutions of the
  free-surface Euler equations that take the form of overtaking
  collisions of unidirectional solitary waves of different amplitude
  on a periodic domain.  As a starting guess, we superpose two Stokes
  waves offset by half the spatial period.  Using an overdetermined
  shooting method, the background radiation generated by collisions of
  the Stokes waves is tuned to be identical before and after each
  collision.  In some cases, the radiation is effectively eliminated
  in this procedure, yielding smooth soliton-like solutions that
  interact elastically forever.  We find examples in which the larger
  wave subsumes the smaller wave each time they collide, and others in
  which the trailing wave bumps into the leading wave, transferring
  energy without fully merging.  Similarities notwithstanding, these
  solutions are found quantitatively to lie outside of the Korteweg-de
  Vries regime. We conclude that quasi-periodic elastic collisions are
  not unique to integrable model water wave equations when the domain
  is periodic.
\end{abstract}

\maketitle

\section{Introduction}

A striking feature of multiple-soliton solutions of integrable model
equations such as the Korteweg-deVries equation, the Benjamin-Ono
equation, and the nonlinear Schr\"odinger equation is that they
interact elastically, leading to time-periodic, relative-periodic, or
quasi-periodic dynamics.  By contrast, the interaction of solitary
waves for the free-surface Euler equations is inelastic.  However, it
has been observed many times in the literature \cite{chan:street:70,
  cooker:97, maxworthy:76, su:mirie, mirie:su,
  zou:su, craig:guyenne:06, milewski:11}
that the residual radiation after a collision of such solitary waves
can be remarkably small.  In the present paper we explore the
possibility of finding nearby time-periodic and relative-periodic
solutions of the Euler equations using a collision of unidirectional
Stokes waves as a starting guess.  Such solutions demonstrate that
recurrent elastic collisions of solitary waves in the spatially
periodic case do not necessarily indicate that the underlying system
is integrable.

A relative-periodic solution is one that returns to a spatial phase
shift of its initial condition at a later time.  This only makes sense
on a periodic domain, where the waves collide repeatedly at regular
intervals in both time and space, with the locations of the collisions
drifting steadily in time.  They are special cases (with $N=2$) of
quasi-periodic solutions, which have the form $u(x,t)=U(\vec\kappa
x+\vec \omega t + \vec\alpha)$ with $U$ an $N$-periodic continuous
function, i.e.~$U\in C\big(\mathbb{T}^N\big)$, and $\vec\kappa$,
$\vec\omega$, $\vec\alpha\in\mathbb{R}^N$.  Throughout the manuscript,
we will use the phrase ``solitary waves'' in a broad sense to describe
waves that, most of the time, remain well-separated from one another
and propagate with nearly constant speed and shape.  ``Stokes waves''
will refer to periodic progressive solutions of the free-surface Euler
equations of permanent form, or waves that began at $t=0$ as a linear
superposition of such traveling waves. They comprise a special class
of solitary waves. ``Solitons'' will refer specifically to
superpositions of $\opn{sech}^2$ solutions of the KdV equation on the
whole line, while ``cnoidal solutions'' will refer to their spatially
periodic, multi-phase counterparts; see \S\ref{sec:kdv} for
elaboration.


It was found in \cite{water2} that decreasing the fluid depth causes
standing waves to transition from large-scale symmetric sloshing
behavior in deep water to pairs of counter-propagating solitary waves
that collide repeatedly in shallow water.  In the present work, we
consider unidirectional waves of different amplitude that collide due
to taller waves moving faster than shorter ones.  We present two
examples of solutions of this type: one where the resulting dynamics
is fully time-periodic; and one where it is relative-periodic,
returning to a spatial phase shift of the initial condition at a later
time. Both examples exhibit behavior typical of collisions of KdV
solitons. In the first, one wave is significantly larger than the
other, and completely subsumes it during the interaction.  In the
second, the waves have similar amplitude, with the trailing wave
bumping into the leading wave and transferring energy without fully
merging.

Despite these similarities, the amplitude of the waves in our examples
are too large for the assumptions in the derivation of the KdV
equation to hold.  In particular, the larger wave in the first example
is more than half the fluid depth in height, and there is significant
vertical motion of the fluid when the waves pass by.  A detailed
comparison of the Euler and KdV equations for waves with these
properties is carried out in \S\ref{sec:kdv}. A review of the
literature on water wave collisions and the accuracy of the KdV model
of water waves is also given in that section.

Rather than compute such solutions by increasing the amplitude from
the linearized regime via numerical continuation, as was done for
counter-propagating waves in \cite{water2}, we use collisions of
right-moving Stokes waves as starting guesses.  The goal is to
minimally ``tune'' the background radiation generated by the Stokes
collisions so that the amount coming out of each collision is
identical to what went into it. In the first example of
\S\ref{sec:num}, we find that the tuned background radiation takes the
form of a train of
traveling waves of smaller wavelength moving to the right more
slowly than either solitary wave.  By contrast, in the
counter-propagating case studied in \cite{water2},
it consists of an array of smaller-wavelength standing waves
oscillating rapidly relative to the time between collisions of the
primary waves.
In the second example of \S\ref{sec:num}, the background radiation is
essentially absent, which is to say that the optimized solution is
free from high-frequency, low-amplitude disturbances in the trough,
and closely resembles a relative-periodic cnoidal solution of KdV.  We
call the collisions in this solution ``elastic'' as they repeat
forever, unchanged up to spatial translation, and there are no
features to distinguish radiation from the waves themselves.  This
process of tuning parameters to minimize or eliminate small-amplitude
oscillations in the wave troughs is reminiscent of Vanden-Broeck's
work \cite{vandenBroeck91} in which oscillations at infinity could be
eliminated from solitary capillary-gravity waves by choosing the
amplitude appropriately.



To search for relative periodic solutions, we use a variant of the
overdetermined shooting method developed by the author and
collaborators in previous work to study several related problems:
time-periodic solutions of the Benjamin-Ono equation
\cite{benj1,benj2} and the vortex sheet with surface tension
\cite{vtxs1,vtxs2}; Hopf bifurcation and stability transitions in
mode-locked lasers \cite{lasers}; cyclic steady-states in rolling
treaded tires \cite{tires1}; self-similarity (or lack thereof) at the
crests of large-amplitude standing water waves \cite{water1}; harmonic
resonance and spontaneous nucleation of new branches of standing water
waves at critical depths \cite{water2}; and three-dimensional standing
water waves \cite{water3d}.  The three approaches developed in these
papers are the adjoint continuation method \cite{benj1,lasers}, a
Newton-Krylov shooting method \cite{tires1}, and a trust region
shooting method \cite{water2} based on the Levenberg-Marquardt
algorithm \cite{nocedal}. We adopt the latter method here to exploit an
opportunity to consolidate the work in computing the Dirichlet-Neumann
operator for many columns of the Jacobian simultaneously, in parallel.

One computational novelty of this work is that we search directly for
large-amplitude solutions of a nonlinear two-point boundary value
problem, without using numerical continuation to get there. This is
generally difficult.
However, in the present case, numerical continuation is also difficult
due to non-smooth bifurcation ``curves'' riddled with Cantor-like gaps
\cite{plotnikov01}, and the long simulation times that occur between
collisions in the unidirectional case. Our shooting method has proven
robust enough to succeed in finding time-periodic solutions, when they
exist, with a poor starting guess. False positives are avoided by
resolving the solutions spectrally to machine accuracy and
overconstraining the minimization problem. Much of the challenge is in
determining the form of the initial condition and the objective
function to avoid wandering off in the wrong direction and falling
into a nonzero local minimum before locking onto a nearby
relative-periodic solution.


\section{Equations of motion}
\label{sec:eqm}

The equations of motion of a free surface $\eta(x,t)$ evolving over
an ideal fluid with velocity potential $\phi(x,y,t)$ may be
written \cite{whitham74,johnson97,craik04,craik05}
\begin{align}
  \label{eq:ww}
    \eta_t &= \phi_y - \eta_x\phi_x, \\[-3pt]
    \notag
    \varphi_t &= P\left[\phi_y\eta_t - \frac{1}{2}\phi_x^2 -
  \frac{1}{2}\phi_y^2 - g\eta\right],
\end{align}
where subscripts denote partial derivatives, $\varphi(x,t) =
\phi(x,\eta(x,t), t)$ is the restriction of $\phi$ to the free
surface, $g=1$ is the acceleration of gravity, $\rho=1$ is the fluid
density, and $P$ is the projection
\begin{equation}
  Pf = f - \frac{1}{2\pi}\int_0^{2\pi} f(x)\,dx,
\end{equation}
where we assume a $2\pi$-periodic domain.  The velocity components
$u=\phi_x$ and $v=\phi_y$ at the free surface can be computed from
$\varphi$ via
\begin{equation}\label{eq:uv:from:G}
  \begin{pmatrix} \phi_x \\ \phi_y \end{pmatrix} =
  \frac{1}{1+\eta'(x)^2}\begin{pmatrix}
    1 & -\eta'(x) \\ \eta'(x) & 1 \end{pmatrix}
  \begin{pmatrix}
    \varphi'(x) \\
    \mc G\varphi(x)
  \end{pmatrix},
\end{equation}
where a prime denotes a derivative and $\mc G$ is the
Dirichlet-Neumann operator \cite{craig:sulem:93}
\begin{equation}\label{eq:DNO:def}
  \mc G\varphi(x)
  = \sqrt{1+\eta'(x)^2}\,\, \der{\phi}{n}(x+i\eta(x))
  = \phi_y - \eta_x\phi_x
\end{equation}
for the Laplace equation, with periodic boundary conditions in $x$,
Dirichlet conditions ($\phi=\varphi$) on the upper boundary, and
Neumann conditions ($\phi_y=0$) on the lower boundary, assumed flat.
We have suppressed $t$ in the notation since time is frozen in the
Laplace equation.  We compute $\mc G\varphi$ using a boundary integral
collocation method \cite{lh76, baker:82, krasny:86, mercer:92, 
baker10} and advance the solution in time using an 8th order
Runge-Kutta scheme \cite{hairer:I} with 36th order filtering
\cite{hou:li:07}. See \cite{water2} for details.



\section{Computation of relative-periodic solutions}
\label{sec:method}

Traveling waves have the symmetry that
\begin{equation}\label{eq:init}
  \eta(x,0) \, \text{ is even}, \qquad \varphi(x,0) \, \text{ is odd.}
\end{equation}
This remains true if $x$ is replaced by $x-\pi$.  As a starting guess
for a new class of time-periodic and relative-periodic solutions, we
have in mind superposing two traveling waves, one centered at $x=0$
and the other at $x=\pi$.  Doing so will preserve the property
(\ref{eq:init}), but the waves will now interact rather than remain
pure traveling waves. A solution will be called \emph{relative
  periodic} if there exists a time $T$ and phase shift $\theta$ such
that
\begin{equation} \label{eq:ts:def}
  \eta(x,t+T) = \eta(x-\theta,t), \qquad\quad
  \varphi(x,t+T) = \varphi(x-\theta,t)
\end{equation}
for all $t$ and $x$.  Time-periodicity is obtained as a special case,
with $\theta\in2\pi\mathbb{Z}$.  We can save a factor of 2 in
computational work by imposing the alternative condition
\begin{equation}
  \label{eq:even:odd}
  \eta(x+\theta/2,T/2) \, \text{ is even}, \qquad\quad
  \varphi(x+\theta/2,T/2) \, \text{ is odd.}
\end{equation}
From this, it follows that
\begin{align*}
  \eta(x+\theta/2,T/2) &= \eta(-x+\theta/2,T/2) = 
  \eta(x-\theta/2,-T/2), \\
  \varphi(x+\theta/2,T/2) &= -\varphi(-x+\theta/2,T/2) = 
  \varphi(x-\theta/2,-T/2).
\end{align*}
But then both sides of each equation in (\ref{eq:ts:def}) agree at
time $t=-T/2$. Thus, (\ref{eq:ts:def}) holds for all time.

In the context of traveling-standing waves in deep water
\cite{trav:stand}, it is natural to define $T$ as twice the value
above, replacing all factors of $T/2$ by $T/4$.  That way a pure
standing wave returns to its original configuration in time $T$
instead of shifting in space by $\pi$ in time $T$.  In the present
work, we consider pairs of solitary waves moving to the right at
different speeds, so it is more natural to define $T$ as the first
(rather than the second) time there exists a $\theta$ such that
(\ref{eq:ts:def}) holds.

\subsection{Objective function}
\label{sec:obj:fun}

We adapt the overdetermined shooting method of \cite{water1,water2} to
compute solutions of (\ref{eq:init})--(\ref{eq:even:odd}).  This
method employs the Levenberg-Marquardt method \cite{nocedal} with
delayed Jacobian updates \cite{water2} to solve the nonlinear least
squares problem described below.

For (\ref{eq:init}), we build the symmetry into the initial
conditions over which the shooting method is allowed to search: we
choose an integer $n$ and consider initial conditions of the form
\begin{equation}\label{eq:init:trav}
  \hat\eta_k(0) = c_{2|k|-1}, \qquad\quad
  \hat\varphi_k(0) = \pm ic_{2|k|},
\end{equation}
where $k\in\{\pm1,\pm2,\dots,\pm \frac{n}{2}\}$ and $\hat\eta_k(t)$,
$\hat\varphi_k(t)$ are the Fourier modes of $\eta(x,t)$,
$\varphi(x,t)$.  The numbers $c_1,\dots,c_n$ are assumed real and all
other Fourier modes (except $\hat\eta_0$) are zero.  We set
$\hat\eta_0$ to the fluid depth so that $y=0$ is a symmetry line
corresponding to the bottom wall.  This is convenient for computing
the Dirichlet-Neumann operator \cite{water2}.  In the formula for
$\hat\varphi_k$, the minus sign is taken if $k<0$ so that
$\hat\varphi_{-k} =\overline{\hat\varphi_k}$.  We also solve
for the period,
\begin{equation}\label{eq:T:theta}
  T=c_{n+1}.
\end{equation}
The phase shift $\theta$ is taken as a prescribed parameter here.
Alternatively, in a study of traveling-standing waves
\cite{trav:stand}, the author defines a traveling parameter $\beta$
and varies $\theta=c_{n+2}$ as part of the algorithm to obtain the
desired value of $\beta$.  This parameter $\beta$ is less meaningful
for solitary wave collisions in shallow water, so we use $\theta$
itself as the traveling parameter in the present study.  We also need
to specify the amplitude of the wave.  This can be done in various
ways, e.g.~by specifying the value of the energy,
\begin{equation*}
  E = \frac{1}{2\pi}\int_0^{2\pi} \jt \frac{1}{2}\varphi\mc{G}\varphi
  + \frac{1}{2}g\eta^2\,dx,
\end{equation*}
by constraining a Fourier mode such as $\hat\eta_1(0)$, or by
specifying the initial height of the wave at $x=0$:
\begin{equation*}
  \eta(0,0) = \hat\eta_0 + \sum_{k=1}^{n/2} 2c_{2k-1}.
\end{equation*}
Thus, to enforce (\ref{eq:even:odd}), we can minimize the objective function
\begin{equation}\label{eq:f}
    f(c) = \frac{1}{2} r(c)^Tr(c),
\end{equation}
where
\begin{align}\label{eq:r:def}
    r_1 = \big(\;\text{choose one:} \quad
    & E-a \quad,\quad
    \hat\eta_1(0)-a \quad,\quad
    \eta(0,0)-a \;\big), \\ \notag
    r_{2j} = \im\{e^{ij\theta/2}\hat\eta_j(T/2)\}, \qquad 
    &r_{2j+1} = \re\{e^{ij\theta/2}\hat\varphi_j(T/2)\}, \qquad
    (1 \le j\le M/2).
\end{align}
Here $a$ is the desired value of the chosen amplitude parameter.
Alternatively, we can impose (\ref{eq:ts:def}) directly by minimizing
\begin{equation}\label{eq:f1}
  \tilde f = \frac{1}{2}r_1^2 + \frac{1}{4\pi}
  \int_0^{2\pi} \left(\big[\eta(x,T)-\eta(x-\theta,0)\big]^2 +
  \big[\varphi(x,T)-\varphi(x-\theta,0)\big]^2\right)dx,
\end{equation}
which also takes the form $\frac{1}{2}r^Tr$ if we define $r_1$ as above and
\begin{equation}
  \label{eq:r1:def}
  \begin{aligned}
    r_{4j-2}+ir_{4j-1} &= \sqrt{2}\left[ \hat\eta_j(T) - e^{-ij\theta}\hat\eta_j(0) \right], \\
    r_{4j}+ir_{4j+1} &= \sqrt{2}\left[ \hat\varphi_j(T) - e^{-ij\theta}\hat\varphi_j(0) \right],
  \end{aligned} \qquad\quad (1\le j\le M/2).
\end{equation}
Note that $f$ measures deviation from evenness and oddness of
$\eta(x+\theta/2,T/2)$ and $\varphi(x+\theta/2,T/2)$, respectively,
while $\tilde f$ measures deviation of $\eta(x+\theta,T)$ and
$\varphi(x+\theta,T)$ from their initial states. In the first example
of \S\ref{sec:num}, we minimize $\tilde f$ directly, while in the second we
minimize $f$ and check that $\tilde f$ is also small, as a means of
validation.  The number of equations, $m=M+1$ for $f$ and $m=2M+1$ for
$\tilde f$, is generally larger than the number of unknowns, $n+1$, due to
zero-padding of the initial conditions. This adds robustness to the
shooting method and causes all Fourier modes varied by the algorithm,
namely those in (\ref{eq:init:trav}), to be well-resolved on the mesh.

\subsection{Computation of the Jacobian}

To compute the $k$th column of the Jacobian $J=\nabla_c r$, which is
needed by the Levenberg-Marquardt method, we solve the linearized
equations along with the nonlinear ones:
\begin{equation}\label{eq:q:qdot}
  \der{}{t}
  \begin{pmatrix} q \\ \dot q \end{pmatrix} =
  \begin{pmatrix} F(q) \\ DF(q)\dot q \end{pmatrix}, \quad
  \begin{aligned}
    q(0) &= q_0 = (\eta_0,\varphi_0), \\
    \dot q(0) &= \dot q_0 = \partial q_0/\partial c_k.
  \end{aligned}
\end{equation}
Here $q=(\eta,\varphi)$, $\dot q=(\dot\eta,\dot\varphi)$, $F(q)$ is
given in (\ref{eq:ww}), $DF$ is its derivative (see \cite{water2} for
explicit formulas), and a dot represents a variational derivative with
respect to perturbation of the initial conditions, not a time
derivative.  To compute $\partial r_i/\partial c_k$ for $i\ge2$ and
$k\le n$, one simply puts a dot over each Fourier mode on the
right-hand side of (\ref{eq:r:def}) or (\ref{eq:r1:def}), including
$\hat\eta_j(0)$ and $\hat\varphi_j(0)$ in (\ref{eq:r1:def}). If
$k=n+1$, then $c_k=T$ and
\begin{equation*}
  \der{r_{2j}}{T} = \im\{e^{ij\theta/2}(1/2)\partial_t\hat\eta_j(T/2)\}, \qquad
  \der{(r_{4j}+ir_{4j+1})}{T} = \sqrt{2}\big[\partial_t\hat\varphi_j(T)\big]
\end{equation*}
in (\ref{eq:r:def}) and (\ref{eq:r1:def}), respectively, with similar
formulas for $\partial(r_{4j-2}+ir_{4j-1})/\partial T$ and $\partial
r_{2j+1}/\partial T$.  The three possibilities for $r_1$ are handled
as follows:
\begin{align*}
  &\text{case 1:} \quad \der{r_1}{c_k} = \dot E
  = \frac{1}{2\pi}\int_0^{2\pi} \left[\dot\varphi\eta_t - \dot\eta\varphi_t \right]_{t=0}dx, 
  \quad (k\le n), \qquad
  \der{r_1}{c_{n+1}} = 0, \\
  &\text{case 2:} \quad \der{r_1}{c_k} = \wdg{\dot\eta}_1(0) = \delta_{k,1},
  \quad (k\le n+1),\\
  &\text{case 3:} \quad \der{r_1}{c_k} = \dot\eta(0,0) = 2\delta_{k,\text{odd}}, \quad
  (k\le n), \qquad \der{r_1}{c_{n+1}} = 0,
\end{align*}
where $\delta_{k,j}$ and $\delta_{k,\text{odd}}$ equal 1 if $k=j$ or $k$ is
odd, respectively, and equal zero otherwise.  The vectors $\dot q$ in
(\ref{eq:q:qdot}) are computed in batches, each initialized with a
different initial perturbation, to consolidate the work in computing
the Dirichlet-Neumann operator during each timestep.  See
\cite{water2,trav:stand} for details.

\section{Numerical results}
\label{sec:num}

As mentioned in the introduction, our idea is to use collisions of
unidirectional Stokes (i.e.~traveling) waves as starting guesses to
find time-periodic and relative periodic solutions of the Euler
equations.  We begin by computing traveling waves of varying wave
height and record their periods. This is easily done in the framework
of \S\ref{sec:method}.  We set $\theta=\pi/64$ (or any other small
number) and minimize $\tilde f$ in (\ref{eq:f1}).  The resulting
``period'' $T$ will give the wave speed via $c=\theta/T$.  Below we
report $T=2\pi c$, i.e.~$T$ is rescaled as if $\theta$ were $2\pi$.
We control the amplitude by specifying
$\hat\eta_1(0)$, which is the second option listed in
\S\ref{sec:method} for defining the first component $r_1$ of the
residual.  A more conventional approach for computing traveling waves
is to substitute $\eta(x-ct)$, $\varphi(x-ct)$ into (\ref{eq:ww}) and
solve the resulting stationary problem (or an equivalent integral
equation) by Newton's method \cite{chen80a, chandler:93,
  milewski:11}. Note that the wave speed $c$ here is unrelated to the
vector $c$ of unknowns in (\ref{eq:init:trav}).


\begin{figure}
\begin{center}
\includegraphics[width=3.3in]{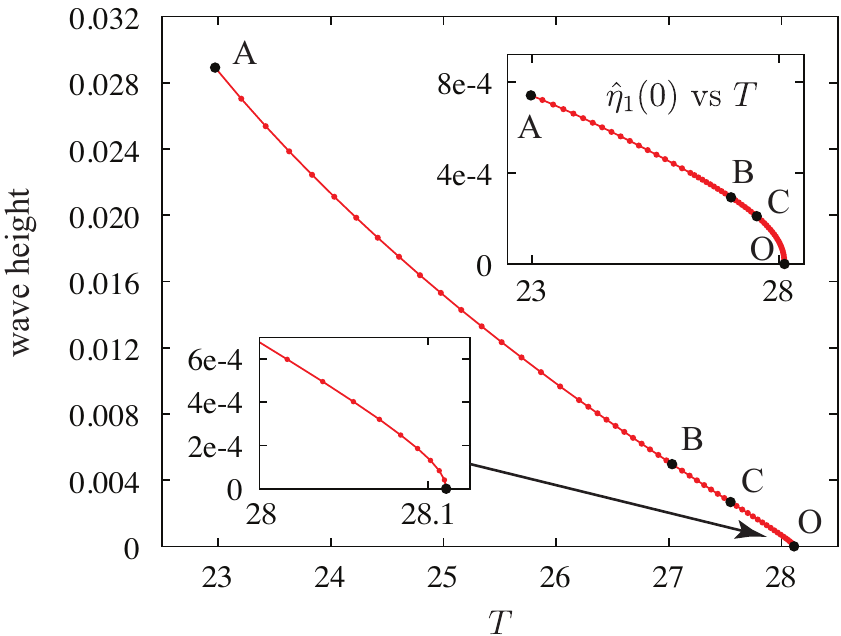}
\end{center}
\caption{\label{fig:bif:stokes} Plots of wave height and first Fourier
  mode versus period for Stokes waves with wavelength $2\pi$ and fluid
  depth $h=0.05$.  The temporal periods are $6T_A=137.843\approx
  137.738 = 5T_C$.}
\end{figure}

\begin{figure}
\begin{center}
\includegraphics[width=\linewidth,trim=5 0 -5 0]{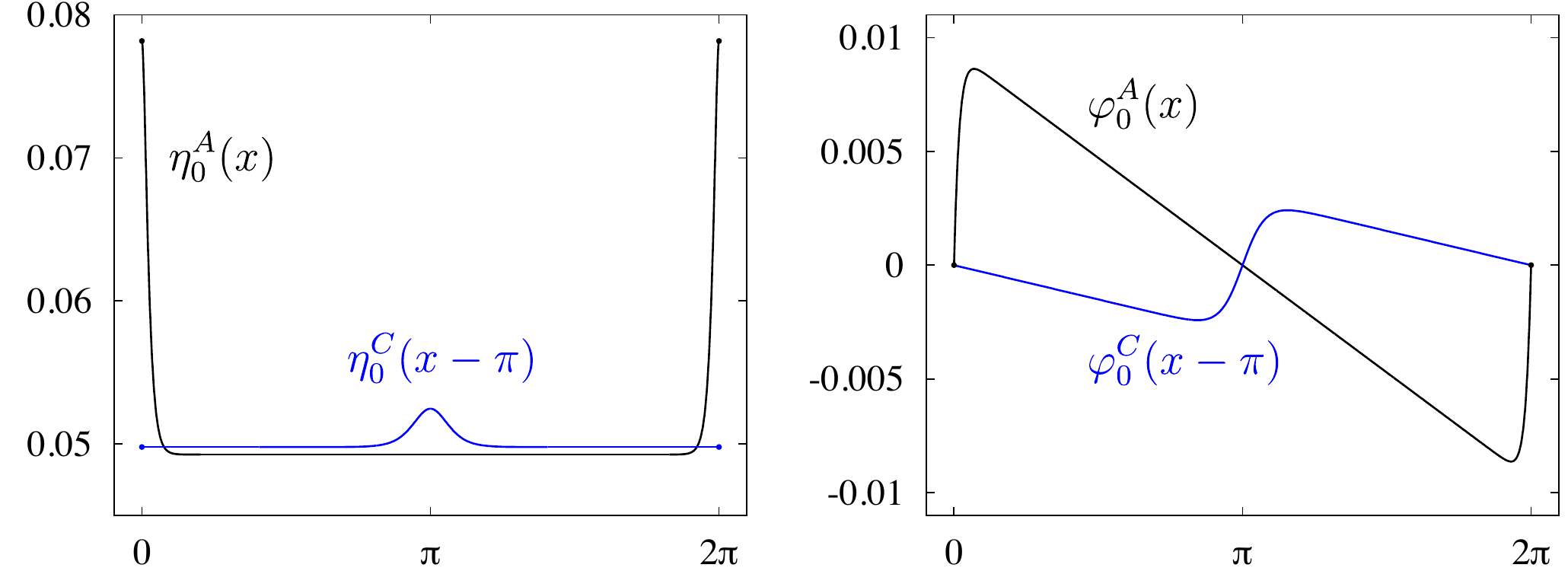}
\end{center}
\caption{\label{fig:AandC} Wave profile and velocity potential of
  Stokes waves labeled A and C in Fig.~\ref{fig:bif:stokes}, plotted
  over one spatial period at $t=0$. }
\end{figure}

\begin{figure}
\begin{center}
\includegraphics[width=\linewidth]{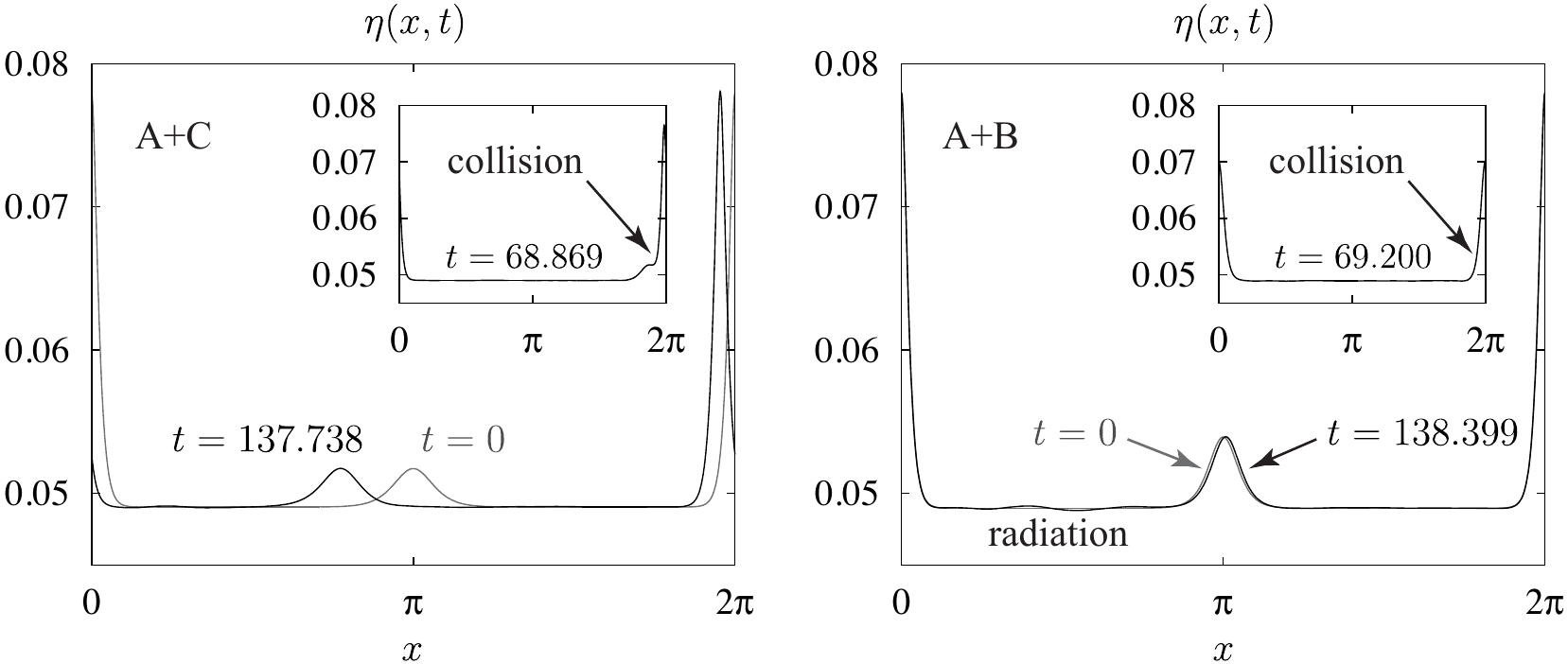}
\end{center}
\caption{\label{fig:align:stokes} 
  Collision of two right-moving Stokes waves that nearly return to
  their initial configuration after the interaction. (left)
  Solutions A and C were combined via (\ref{eq:AandC}) and evolved
  through one collision to $t=137.738$.  (right) Through trial and
  error, we adjusted the amplitude of the smaller Stokes wave and the
  simulation time to obtain a nearly time-periodic solution. }
\end{figure}

\begin{figure}
\begin{center}
\includegraphics[width=\linewidth]{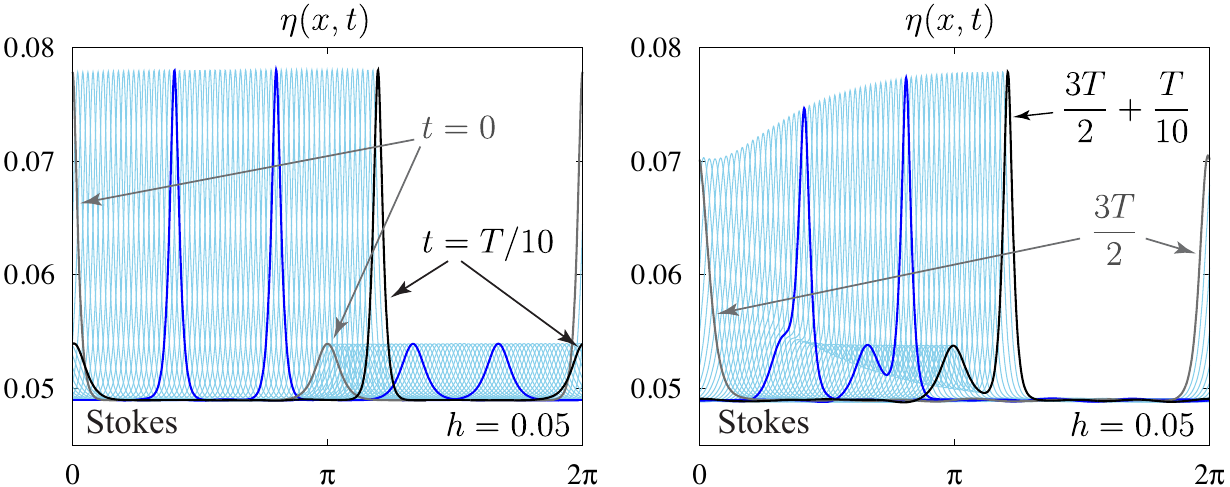}
\end{center}
\caption{\label{fig:evol:kdv1} Evolution of two Stokes waves that
  collide repeatedly, at times $t\approx T/2+kT$, $k\ge0$. (left)
  Traveling solutions A and B in Figure~\ref{fig:bif:stokes} were
  initialized with wave crests at $x=0$ and $x=\pi$, respectively. The
  combined solution is approximately time-periodic, with period
  $T=138.399$.  (right) The same solution, at later times, starting
  with the second collision ($t=3T/2$).}
\end{figure}

\begin{figure}
\begin{center}
\quad\;\;\includegraphics[height=2in]{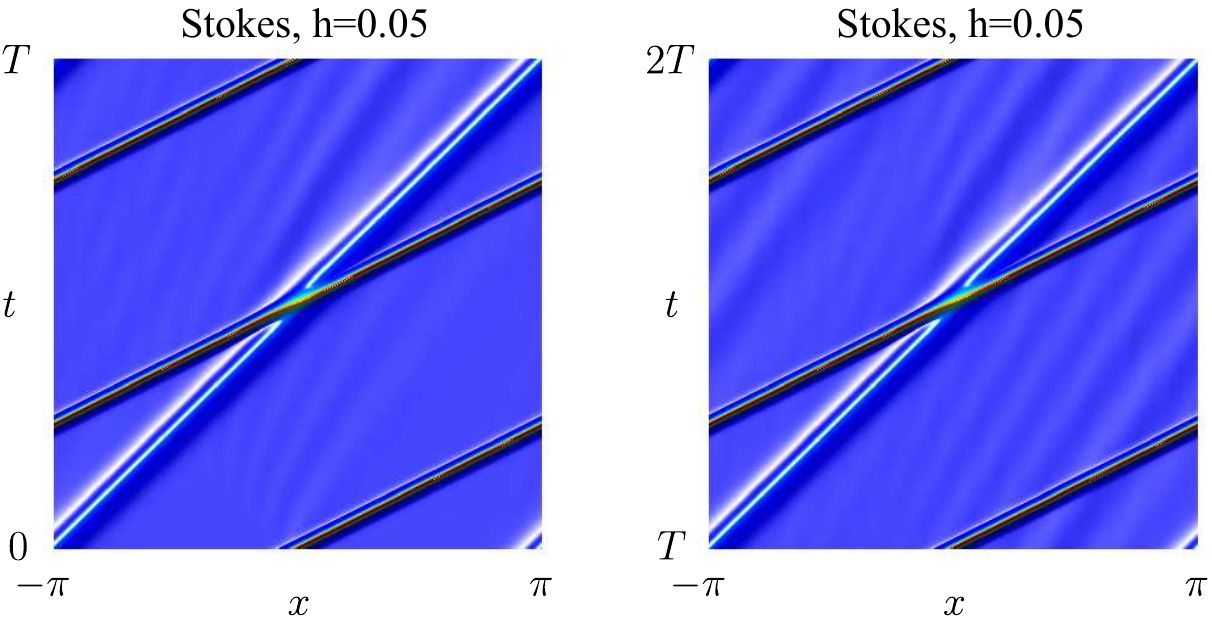}
\end{center}
\caption{\label{fig:pov:kdv1} A different view of the solutions in
  Figure~\ref{fig:evol:kdv1} shows the generation of background
  waves. Shown here are the functions $\eta(x+8\pi t/T,t)$, which give
  the dynamics in a frame moving to the right fast enough to traverse
  the domain four times in time $T$.  In a stationary frame, the
  smaller and larger solitary waves traverse the domain 5 and 6 times,
  respectively.}
\end{figure}

With traveling waves in hand, out next goal is to collide two of them
and search for a nearby time-periodic solution, with $\theta=0$.  As
shown in Figure~\ref{fig:bif:stokes}, varying $\hat\eta_1(0)$ from 0
to $7.4\times 10^{-4}$ causes the period of a Stokes wave with
wavelength $\lambda=2\pi$ and mean fluid depth $h=0.05$ to decrease
from $T_O=28.1110$ to $T_A=22.9739$, and the wave height (vertical
crest-to-trough distance) to increase from 0 to $0.02892$.  Solution C
is the closest among the Stokes waves we computed to satisfying
$5T_C=6T_A$, where $p=5$ is the smallest integer satisfying
$\frac{p+1}{p}T_A<T_O$.  We then combine solution A with a spatial
phase shift of solution C at $t=0$.  The resulting initial conditions
are
\begin{equation}\label{eq:AandC}
  \begin{aligned}
  \eta^{A+C}_0(x) &= h + \big[\eta^A_0(x)-h\big] + \big[\eta^C_0(x-\pi)-h\big], \\
  \varphi^{A+C}_0(x) &= \varphi^A_0(x) + \varphi^C_0(x-\pi),
  \end{aligned}
\end{equation}
where $h=0.05$ is the mean fluid depth.  Plots of $\eta_0^A(x)$,
$\eta_0^C(x-\pi)$, $\varphi_0^A(x)$ and $\varphi_0^C(x-\pi)$
are shown in Figure~\ref{fig:AandC}.  If the waves did not interact,
the combined solution would be time-periodic (to the extent that
$5T_C=6T_A$, i.e.~to about $0.076\%$).  But the waves do interact. In
addition to the complicated interaction that occurs when they collide,
each slows the other down between collisions by introducing a negative
gradient in the velocity potential between its own wave
crests. Indeed, as shown in the right panel of Figure~\ref{fig:AandC},
the velocity potential increases rapidly across a right-moving
solitary wave and decreases elsewhere to achieve spatial
periodicity. The decreasing velocity potential induces a background
flow opposite to the direction of travel of the other wave. In the
left panel of Figure~\ref{fig:align:stokes}, we see that the net
effect is that neither of the superposed waves has returned to its
starting position at $t=5T_C$, and the smaller wave has experienced a
greater net decrease in speed. However, as shown in the right panel,
by adjusting the amplitude of the smaller wave (replacing solution C
by B) and increasing $T$ slightly to $138.399$, we are able to
hand-tune the Stokes waves to achieve $\tilde
f\approx5.5\times10^{-8}$, where $\theta$ is set to zero in
(\ref{eq:f1}).
Note that as $t$ varies from 0 to $T/10$ in the left panel of
Figure~\ref{fig:evol:kdv1}, the small wave advances by $\pi$ units to
the right while the large wave advances by $1.2\pi$ units.  The waves
collide at $t=T/2$.  This generates a small amount of radiation, which
can be seen at $t=T$ in the right panel of
Figure~\ref{fig:align:stokes}. Some radiation behind the large
wave is present for all $t>0$, as shown in Figure~\ref{fig:pov:kdv1}.

\begin{figure}
\begin{center}
\includegraphics[width=\linewidth]{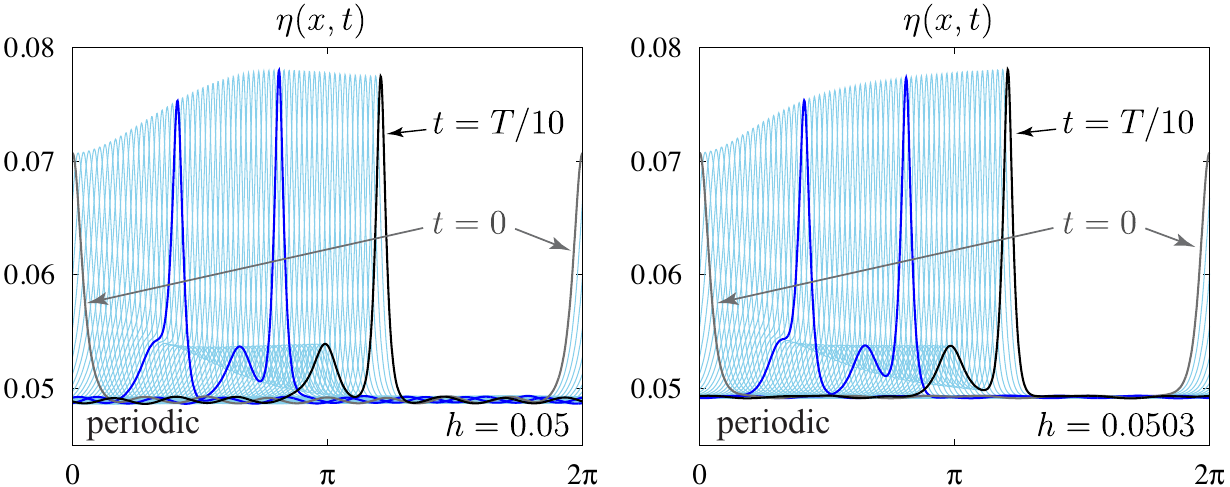}
\end{center}
\caption{\label{fig:evol:kdv2} Time-periodic solutions near the Stokes
  waves of Figure~\ref{fig:evol:kdv1}. (left) $h=0.05$, $\eta(0,0) =
  0.0707148$, $T=138.387$, $\tilde f=4.26\times10^{-27}$. (right)
  $h=0.0503$, $\eta(0,0)=0.0707637$, $T=138.396$, $\tilde f=1.27\times
  10^{-26}$.  The background radiation was minimized by hand in the
  right panel by varying $h$ and $\eta(0,0)$.}
\end{figure}

Before minimizing $\tilde f$, we advance the two Stokes waves to the time
of the first collision, $t=T/2$. At this time, the larger solitary
wave has traversed the domain 3 times and the smaller one 2.5 times,
so their peaks lie on top of each other at $x=0$.  The reason to do
this is that when the waves merge, the combined wave is shorter,
wider, and smoother than at any other time during the evolution.
Quantitatively, the Fourier modes of $\hat\eta_k(t)$ and
$\hat\varphi_k(t)$ decay below $10^{-15}$ for $k\ge600$ at $t=0$, and
$k\ge200$ when $t=T/2$.  Thus, the number of columns needed in the
Jacobian is reduced by a factor of 3, and the problem becomes more
overdetermined, hence more robust.  For the calculation of a
time-periodic solution, we let $t=0$ correspond to this merged state,
which affects the time labels when comparing
Figures~\ref{fig:evol:kdv1} and~\ref{fig:evol:kdv2}.  As a final
initialization step, we project onto the space of initial conditions
satisfying (\ref{eq:init:trav}) by zeroing out the imaginary parts of
$\hat\eta_k(0)$ and the real parts of $\hat\varphi_k(0)$, which are
already small. Surprisingly, this improves the time-periodicity of the
initial guess in (\ref{eq:f1}) to $\tilde f = 2.3\times 10^{-8}$.

We emphasize that our goal is to find \emph{any} nearby time-periodic
solution by adjusting the initial conditions to drive $\tilde f$ to
zero.  Energy will be conserved as the solution evolves from a given
initial condition, but is only imposed as a constraint (in the form of
a penalty) on the search for initial conditions when the first
component of the residual in (\ref{eq:r:def}) is set to $r_1=E-a$. In
the present calculation, we use $r_1=\eta(0,0)-a$ instead. In the
second example, presented below, we will constrain energy.  In either
case, projecting onto the space of initial conditions satisfying
(\ref{eq:init:trav}) can cause $r_1$ to increase, but it will decrease
to zero in the course of minimizing $\tilde f$. This projection is
essential for the symmetry arguments of \S\ref{sec:obj:fun} to work.

\begin{figure}
\begin{center}
\includegraphics[height=2in]{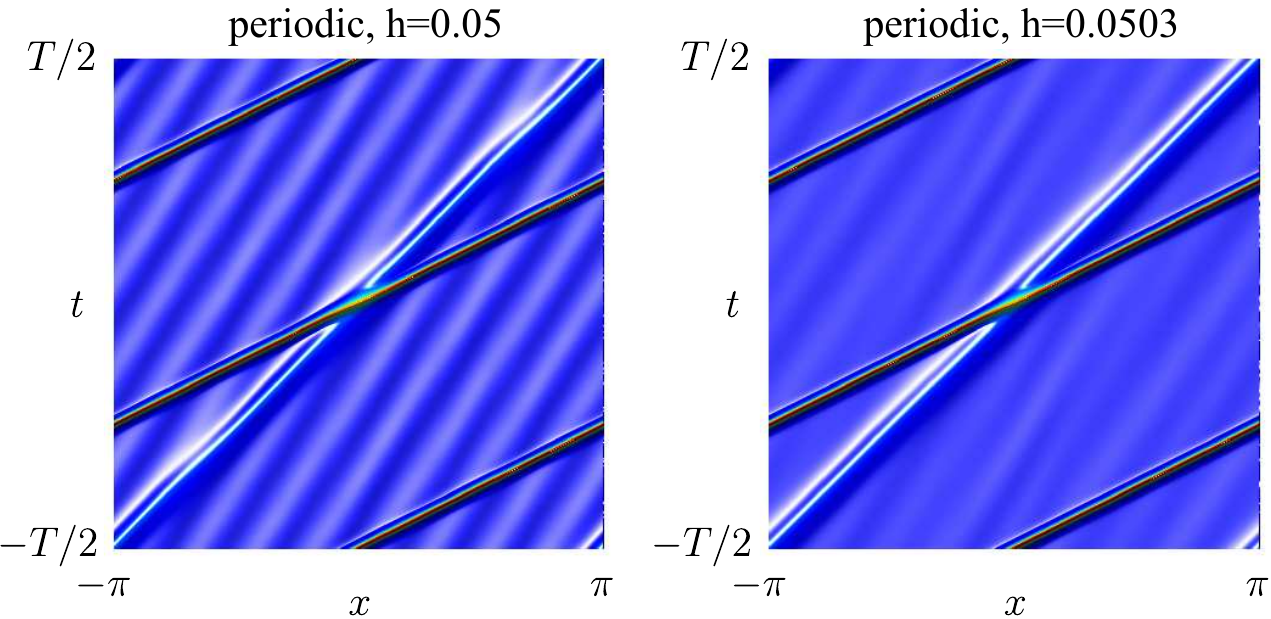}
\end{center}
\caption{\label{fig:pov:kdv2} Same as Figure~\ref{fig:pov:kdv1}, but
  showing the time-periodic solutions of Figure~\ref{fig:evol:kdv2}
  instead of the Stokes waves of Figure~\ref{fig:evol:kdv1}.  The
  Stokes waves generate new background radiation with each collision
  while the time-periodic solutions are synchronized with the
  background waves to avoid generating additional disturbances. }
\end{figure}

We minimize $\tilde f$ subject to the constraint
$\eta(0,0)=0.0707148$, the third case described in \S\ref{sec:method}
for specifying the amplitude. This causes $\tilde f$ to decrease from
$2.3\times 10^{-8}$ to $4.26\times 10^{-27}$ using $M=1200$ grid
points and $N=1200$ time-steps (to $t=T$).  The results are shown in
the left panel of Figures~\ref{fig:evol:kdv2} and~\ref{fig:pov:kdv2}.
The main difference between the Stokes collision and this nearby
time-periodic solution is that the Stokes waves generate additional
background ripples each time they collide while the time-periodic
solution contains an equilibrium background wave configuration that
does not grow in amplitude after the collision.  While the background
waves in the counter-propagating case (studied in \cite{water2}) look
like small-amplitude standing waves, these background waves travel to
the right, but slower than either solitary wave.  After computing the
$h=0.05$ time-periodic solution, we computed 10 other solutions with
nearby values of $h$ and $\eta(0,0)$ to try to decrease the amplitude
of the background radiation. The best solution we found (in the sense
of small background radiation) is shown in the right panel of
Figures~\ref{fig:evol:kdv2} and~\ref{fig:pov:kdv2}, with $h=0.0503$
and $\eta(0,0)=0.0707637$. The amplitude of the background waves of
this solution are comparable to that of the Stokes waves after two
collisions.


Our second example is a relative periodic solution in which the
initial Stokes waves (the starting guess) are B and C in
Figure~\ref{fig:bif:stokes} instead of A and C.  As before, solution C
is shifted by $\pi$ when the waves are combined initially, just as in
(\ref{eq:AandC}).  Because the amplitude of the larger wave has been
reduced, the difference in wave speeds is smaller, and it takes much
longer for the waves to collide.  If the waves did not interact, we
would have
\begin{equation}\label{eq:cBcC}
  c_{B,0} = 0.23246089, \quad c_{C,0} = 0.22808499, \quad
  T_0 = \frac{2\pi}{c_{B,0}-c_{A,0}} = 1435.86,
\end{equation}
where wave B crosses the domain $53.1230$ times in time $T_0$ while
wave C crosses the domain $52.1230$ times.  The subscript 0 indicates
that the waves are assumed not to interact. Since the waves do
interact, we have to evolve the solution numerically to obtain useful
estimates of $T$ and $\theta$.  We arbitrarily rounded $T_0$ to 1436
and made plots of the solution at times $\Delta t = T_0/1200$.  We
found that $\eta$ is nearly even (up to a spatial phase shift) for the
first time at $463\Delta t=554.057$.  This was our initial guess for
$T/2$.  The phase shift required to make $\eta(x+\theta/2,T/2)$
approximately even and $\varphi(x+\theta/2,T/2)$ approximately odd was
found by graphically solving $\varphi(x,T/2)=0$. This gives the
initial guess $\theta/2=2.54258$.  This choice of $T$ and $\theta$
(with $\eta^{B+C}$ and $\varphi^{B+C}$ as initial conditions) yields
$f=2.0\times10^{-11}$ and $\tilde f=1.5\times10^{-10}$.  We then minimize
$f$ holding $E$ and $\theta$ constant, which gives
$f=2.1\times10^{-29}$ and $\tilde f=3.0\times10^{-26}$.
We note that $\tilde f$ is computed over $[0,T]$, twice the
time over which the solution was optimized by minimizing $f$,
and provides independent confirmation of the accuracy of the solution
and the symmetry arguments of \S\ref{sec:obj:fun}.

\begin{figure}
\begin{center}
\includegraphics[width=\linewidth]{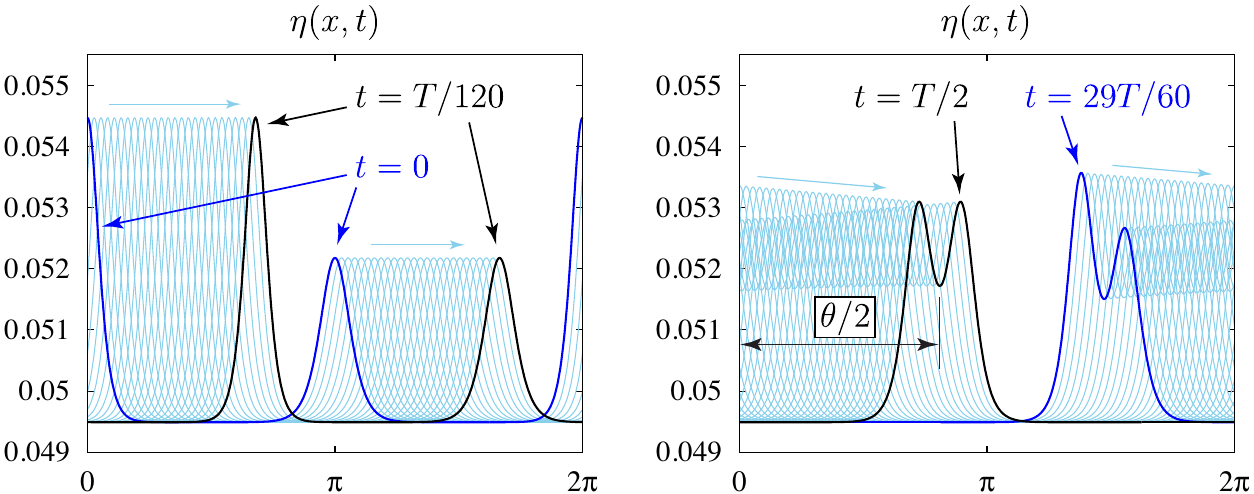}
\end{center}
\caption{\label{fig:evol:kdv3} A relative-periodic solution found
  using a superposition of the Stokes waves labeled B and C in
  Figure~\ref{fig:bif:stokes} as a starting guess.  Unlike the
  previous case, the waves do not fully merge at $t=T/2$. }
\end{figure}

The results are plotted in Figure~\ref{fig:evol:kdv3}.  We omit a
plot of the initial guess (the collision of Stokes waves) as it is
indistinguishable from the minimized solution.  In fact, the
relative change in the wave profile and velocity potential is
about $0.35$ percent,
\begin{equation}
  \left(\frac{
  \|\eta_\text{Stokes} - \eta_\text{periodic}\|^2 +
  \|\varphi_\text{Stokes} - \varphi_\text{periodic}\|^2}{
  \|\eta_\text{Stokes} - h\|^2 + \|\varphi_\text{Stokes}\|^2}
  \right)^{1/2} \le 0.0035,
\end{equation}
and $T/2$ changes even less, from 554.057 (Stokes) to 554.053
(periodic). By construction, $E$ and $\theta/2$ do not change at all.
It was not necessary to evolve the Stokes waves to $T/2$, shift space
by $\theta/2$, zero out Fourier modes that violate the symmetry
condition (\ref{eq:init}), and reset $t=0$ to correspond to this new
initial state.  Doing so increases the decay rate of the Fourier modes
(slope of $\ln|\hat\eta_k|$ vs $k$) by a factor of 1.24 in this
example, compared to 3.36 in the previous example, where it is
definitely worthwhile.

The large change from $T_0/2 = 717.93$ to $T/2=554.053$ is due to
nonlinear interaction of the waves.  There are two main factors
contributing to this change in period.  The first is that the waves do
not fully combine when they collide. Instead, the trailing wave runs
into the leading wave, passing on much of its amplitude and speed.
The peaks remain separated by a distance of $d=0.52462$ at $t=T/2$,
the transition point where the waves have the same amplitude. Thus,
the peak separation changes by $\pi-d$ rather than $\pi$ in half a
period.  The second effect is that the larger wave slows down the
smaller wave more than the smaller slows the larger.  Recall from
Fig.~\ref{fig:AandC} that each wave induces a negative potential
gradient across the other wave that generates a background flow
opposing its direction of travel.  Quantitatively, when the waves are
well separated, we find that the taller and smaller waves travel at
speeds $c_B=0.231077=0.994049c_{B,0}$ and
$c_C=0.226153=0.991531c_{C,0}$, respectively.  The relative speed is
then $(c_B-c_C) = 1.12526(c_{B,0}-c_{C,0})$.  Thus,
\begin{equation}\label{eq:ineq}
  \frac{\pi-d}{c_B-c_C}  < \frac{T}{2} <
  \frac{\pi-d}{c_{B,0}-c_{C,0}} < \frac{T_0}{2} = \frac{\pi}{c_{B,0}-c_{C,0}},
\end{equation}
with numerical values $531.5<554.1<598.0<717.9$.
This means that both effects together have overestimated the
correction needed to obtain $T$ from $T_0$.  This is because the
relative speed slows down as the waves approach each other, which is
expected since the amplitude of the trailing wave decreases and the
amplitude of the leading wave increases in this interaction
regime. Indeed, the average speed of the waves is
\begin{equation}\label{eq:average:speed}
  \overline{c_B} = \frac{\theta/2 - d/2}{T/2} = 
  0.993388c_{B,0}, \qquad
  \overline{c_C} = \frac{\theta/2 + d/2 - \pi}{T/2} = 
  0.991737c_{C,0},
\end{equation}
which are slightly smaller and larger, respectively, than their speeds
when well separated. Note that $T/2$ in (\ref{eq:ineq}) may be written
$T/2=(\pi-d)/(\overline{c_B} - \overline{c_C})$.
We used $\theta/2=2.54258+40\pi$ in (\ref{eq:average:speed}) to
account for the 20 times the waves cross the domain $(0,2\pi)$ in time
$T/2$ in addition to the offset shown in Figure~\ref{fig:evol:kdv3}.

\section{Comparison with KdV}
\label{sec:kdv}

In the previous section, we observed two types of overtaking
collisions for the water wave: one in which the larger wave
completely subsumes the smaller wave for a time, and one where the two
waves remain distinct throughout the interaction.  Similar behavior
has of course been observed for the Korteweg-de Vries equation, which
was part of our motivation for looking for such solutions.  Lax
\cite{lax:1968} classified overtaking collisions of two KdV solitons
as bimodal, mixed, or unimodal. Unimodal and bimodal waves are
analogous to the ones we computed above, while mixed mode collisions
have the larger wave mostly subsume the smaller wave at the beginning
and end of the interaction, but with a two-peaked structure
re-emerging midway through the interaction.  Lax showed that if
$1<c_1/c_2<A=(3+\sqrt{5})/2$, the collision is bimodal; if
$c_1/c_2>3$, the collision is unimodal; and if $A<c_1/c_2<3$, the
collision is mixed.  Here $c_1$ and $c_2$ are the wave speeds of the
trailing and leading waves, respectively, at $t=-\infty$.  Leveque
\cite{leveque:87} has studied the asymptotic dynamics of the
interaction of two solitons of nearly equal amplitude.  Zou and Su
\cite{zou:su} performed a computational study of overtaking water wave
collisions, compared the results to KdV interactions, and found that
the water wave collisions ceased to be elastic at third order. Craig
\emph{et.~al.}~\cite{craig:guyenne:06} also found that solitary water
waves collide inelastically. This does not conflict with our results
since we optimize the initial conditions to make the collision
elastic.  Head on collisions have been studied numerically by Su and
Mirie \cite{su:mirie,mirie:su}, experimentally by Maxworthy
\cite{maxworthy:76}, and by a mixture of analysis and computation by
Craig \emph{et.~al.} \cite{craig:guyenne:06}.

Validation of KdV as a model of water waves has also been studied
extensively. A formal derivation may be found in Ablowitz and Segur
\cite{ablowitz:segur}. Rigorous justification has been given by Bona,
Colin and Lannes \cite{bona:lannes}, building on earlier work by Craig
\cite{craig:kdv} as well as Schneider and Wayne
\cite{schneider:wayne}.  According to \cite{bona:lannes}, some gaps
still exist in the theory in the spatially periodic case.
Experimental studies of the validity of KdV for describing surface
waves have been performed by Zabusky and Galvin \cite{zabusky:galvin}
as well as Hammack and Segur \cite{hammack:segur:74}.  Recently,
Ostrovsky and Stepanyants \cite{ostrovsky} have compared internal
solitary waves in laboratory experiments to the predictions of various
model equations, including KdV, and give a comprehensive overview of
the literature on this subject \cite{ostrovsky}.

Our objective in this section is to determine quantitatively whether
the solutions of the water wave equations that we computed in
\S\ref{sec:num} are well-approximated by the KdV equation.
Following Ablowitz and Segur \cite{ablowitz:segur}, we
introduce a small parameter $\veps$ and dimensionless variables
\begin{equation*}
  \hat y = \frac{y}{h}, \qquad
  \hat x = \sqrt{\veps}\frac{x}{h}, \qquad
  \hat t = \sqrt{\frac{\veps g}{h}} t, \qquad
  \hat \eta = \frac{\eta}{\veps h}, \qquad
  \hat \phi = \frac{\phi}{\sqrt{\veps g h^3}},
\end{equation*}
where $h$ is the fluid depth.  We assume the bottom boundary is at
$y=-h$ rather than 0 in this derivation, so that $\hat y$ runs from $-1$
to $\veps\hat\eta$.  The Laplacian becomes $\Delta_\veps = h^{-2}\big(
\veps\partial_{\hat{x}}^2 + \partial_{\hat{y}}^2 \big)$, which allows
for $\hat\phi = \hat\phi_0 + \veps\hat\phi_1 + \veps^2\hat\phi_2 +
\cdots$ to be computed order by order, with leading terms satisfying
\begin{equation*}
  \hat\phi_{0,\hat y} = 0, \qquad
  \hat\phi_1 = -\frac{1}{2}(1+\hat y)^2\hat\phi_{0,\hat x\hat x}, \qquad
  \hat\phi_2 = \frac{1}{24}(1+\hat y)^4\hat\phi_{0,\hat x\hat x\hat x\hat x}.
\end{equation*}
Here we used $\Delta\phi=0$ and $\phi_y(x,-h)=0$.  Note that
$\hat\phi_0$ is independent of $\hat y$, and agrees with the velocity
potential $\phi$ on the bottom boundary, up to rescaling:
\begin{equation*}
  \hat\phi_0(\hat x,\hat t) = (\veps gh^3)^{-1/2}\phi(x,-h,t).
\end{equation*}
From the equations of motion, $\eta_t = \phi_y - \eta_x\phi_x$ and
$\phi_t + \frac{1}{2}\phi_x^2 + \frac{1}{2}\phi_y^2 + g\eta = 0$, one finds
that
\begin{align*}
  \hat\eta_{\hat t} + \hat u_{\hat x} &=
  \veps\big\{ \jt \frac{1}{6}\hat u_{\hat x\hat x\hat x} - (\hat\eta\hat u)_{\hat x}
  \big\} + O(\veps^2), \\
  \hat u_{\hat t} + \hat\eta_{\hat x} &= \veps \big\{ \jt
  \frac{1}{2} \hat u_{\hat x\hat x\hat t} - \frac{1}{2}\partial_{\hat x}(\hat u)^2
  \big\} + O(\veps^2),
\end{align*}
where $\hat u(\hat x,\hat t) = \partial_{\hat x}\hat\phi_0(\hat x,\hat t)$.
Expanding $\hat\eta=\hat\eta_0 + \veps\hat\eta_1 +\cdots$,
$\hat u=\hat u_0 + \veps\hat u_1 +\cdots$, we find that
\begin{equation*}
  \begin{aligned}
    \hat\eta_0 = f(\hat x - \hat t; \tau) + g(\hat x + \hat t; \tau), \\
    \hat u_0 = f(\hat x - \hat t; \tau) - g(\hat x + \hat t; \tau),
  \end{aligned} \qquad
  \begin{aligned}
    2f_\tau + 3ff_r + (1/3)f_{rrr} &= 0, \\
    -2g_\tau + 3gg_l + (1/3)g_{lll} &= 0,
  \end{aligned}
\end{equation*}
where we have introduced characteristic coordinates $r=\hat x-\hat t$,
$l = \hat x + \hat t$ as well as a slow time scale $\tau=\veps\hat t$
to eliminate secular growth in the solution with respect to $r$ and
$l$ at first order in $\veps$; see \cite{ablowitz:segur} for details.
The notational conflict of $g(l,\tau)$ with the acceleration of
gravity, $g$, is standard, and will not pose difficulty below.


\begin{figure}
\begin{center}
\includegraphics[width=\linewidth,trim=0 12 0 0]{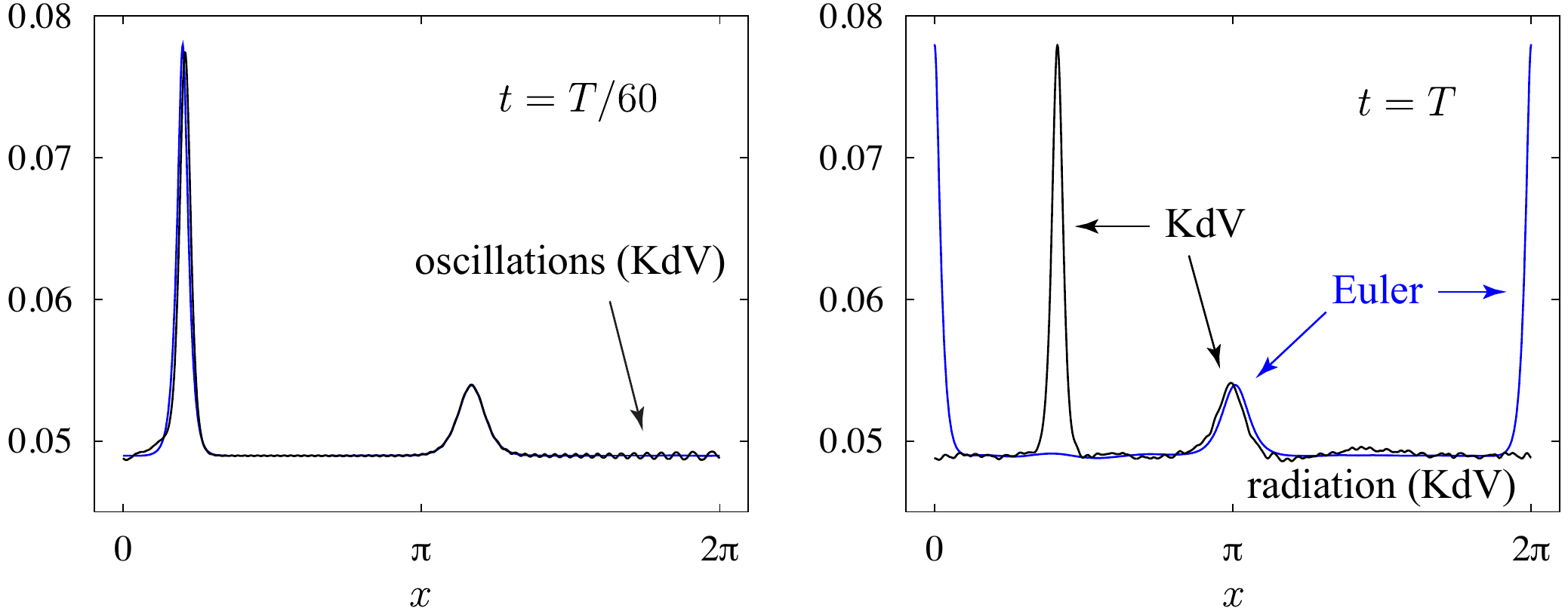}
\end{center}
\caption{\label{fig:kdv:cmp1} Comparison of the solutions of the KdV
  and Euler equations, initialized identically with the superposition
  of Stokes waves labeled A and B in Figure~\ref{fig:bif:stokes}. The
  final time $T$ is set to $138.399$, as in
  Fig.~\ref{fig:align:stokes}, when the Euler solution nearly returns
  to its initial configuration after a single overtaking collision.  }
\end{figure}

In our case, the waves travel to the right, so we may set
$g(l,\tau)=0$ in the formulas above.  Returning to dimensional
variables, we then have
\begin{equation*}
  \eta(x,t) = h\veps f\left(\sqrt\veps\left(\frac{x}{h} - \sqrt{\frac{g}{h}} t
      \right),\sqrt{\frac{g}{h}}\veps^{3/2}t\right),
\end{equation*}
which satisfies
\begin{equation}\label{eq:dim:kdv}
  \eta_t + \alpha \eta_x + \frac{3\sqrt{gh}}{2h}\eta\eta_x +
  \frac{1}{6}\sqrt{gh}\,h^2\eta_{xxx} = 0,
\end{equation}
where $\alpha=\sqrt{gh}$.  Note that $\veps$ drops out. For comparison
with the results of \S\ref{sec:num}, we will add $h$ to $\eta$ and set
$\alpha=-\frac{1}{2}\sqrt{gh}$ instead.  In Figure~\ref{fig:kdv:cmp1},
we compare the solution of (\ref{eq:dim:kdv}), with initial condition
$\eta(x,0) = \eta_0^{A+B}(x)$, defined similarly to $\eta_0^{A+C}(x)$
in (\ref{eq:AandC}), to the solution of the free-surface Euler
equations shown in Figs.~\ref{fig:align:stokes}
and~\ref{fig:evol:kdv1}.  Shortly after the waves are set in motion,
the KdV solution develops high-frequency oscillations behind the
larger peak that travel left and quickly fill up the computational
domain with radiation. The solution of the Euler equations remains
much smoother.  The large peak of the KdV solution also travels
$3.4\%$ faster, on average, than the corresponding peak of the Euler
solution, so that at $t=138.399$, when the taller Euler wave has
traversed the domain 6 times, the taller KdV wave has traversed it
$6.206$ times. For our purposes, these discrepancies are much too
large for KdV to be a useful model, and we conclude that the first
example in \S\ref{sec:num} is well outside of the KdV regime.

In this comparison, timestepping the KdV equation was done with the 8
stage, 5th order implicit/explicit Runge-Kutta method of Kennedy and
Carpenter \cite{carpenter}. Spatial derivatives were computed
spectrally using the 36th order filter of Hou and Li \cite{hou:li:07}.
We found that 2048 spatial grid points and 96000 timesteps was
sufficient to reduce the error at $t=138.399$ below $5\times 10^{-6}$
near the larger peak and below $6\times 10^{-7}$ elsewhere, based on
comparing the solution to one with 3072 grid points and 192000
timesteps. Our solutions of the Euler equations are much more accurate
since there are no second or third spatial derivative terms present to
make the equations stiff.  Thus, we can use 8th or 15th order explicit
timestepping rather than 5th order implicit/explicit
timestepping. Monitoring energy conservation and performing mesh
refinement studies suggests that we obtain 13--14 digits of accuracy
in the solutions of the Euler equations, at which point roundoff error
prevents further improvement in double-precision arithmetic.

\begin{figure}
\begin{center}
\includegraphics[width=\linewidth]{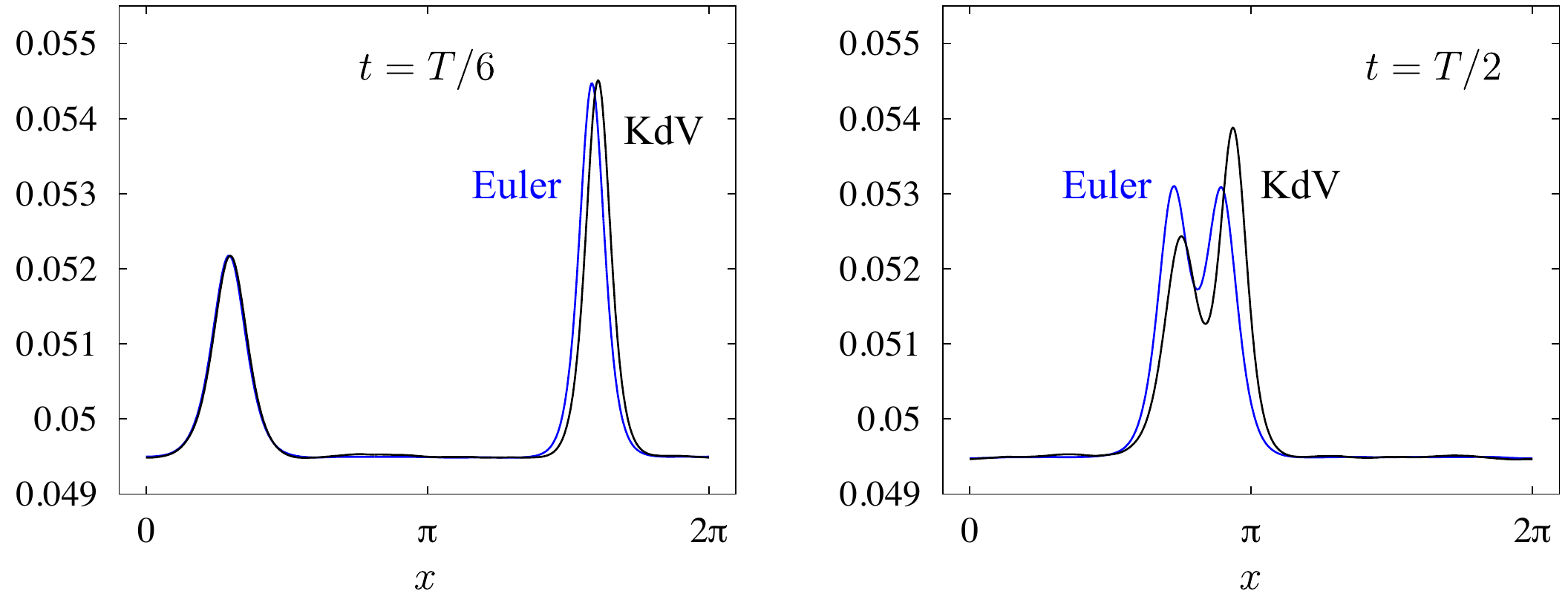}
\end{center}
\caption{\label{fig:kdv:cmp2} Comparison of the solutions of the KdV
  and Euler equations, both initialized with the superposition
  of Stokes waves labeled B and C in Figure~\ref{fig:bif:stokes}.
  $T=1108.11$ here.  }
\end{figure}

In Figure~\ref{fig:kdv:cmp2}, we repeat this computation using initial
conditions corresponding to the superposition of Stokes waves
$\eta_0^{B+C}(x)$, which was used as a starting guess for the
second example of
\S\ref{sec:num}. This time the KdV solution does not develop visible
high-frequency radiation in the wave troughs, and the solutions of KdV
and Euler remain close to each other for much longer.  However, the
interaction time for a collision also increases, from $T=138.399$ in
the first example to $T=1108.11$ here.  In Fig.~\ref{fig:kdv:cmp2}, by
$t=T/6$, the taller KdV and Euler waves have visibly separated from
each other, and by $t=T/2$, when the Euler waves have reached their
minimum approach distance, the KdV solution is well ahead of the Euler
solution.  Thus, while there is good qualitative agreement between the
KdV and Euler solutions, they do not agree quantitatively over the
time interval of interest. From this point of view, the second example
of \S\ref{sec:num} also lies outside of the KdV regime.

An alternative measure of the agreement between KdV and Euler is to
compare the solutions from \S\ref{sec:num} with nearby
relative-periodic solutions of KdV. In other words, we wish to
quantify how much the initial conditions and period have to be
perturbed to convert a relative-periodic solution of the Euler
equations into one for the KdV equations.  Since we used a
superposition of Stokes waves for the initial guess to find
time-periodic and relative-periodic solutions of the Euler equations,
we will use a similar superposition (of cnoidal waves) for KdV.  The
vertical crest-to-trough heights of the three Stokes waves considered
in \S\ref{sec:num} are
\begin{equation}\label{eq:H:ABC}
  H_A = 0.028918699, \qquad
  H_B = 0.004973240, \qquad
  H_C = 0.002683648.
\end{equation}
Well-known \cite{kdv:1895,dingemans} periodic traveling wave solutions 
of (\ref{eq:dim:kdv}) are given by
\begin{gather*}
  \eta(x,t) = h - H + \frac{H}{m}\left(1 - \frac{E(m)}{K(m)}\right) + H\opn{cn}^2\left(
    2K(m)\frac{x-ct}{\lambda}\bigg\vert m\right), \\
  \lambda = \sqrt{\frac{16mh^3}{3H}}\,K(m), \qquad
  c = \left[1 - \frac{H}{2h} + \frac{H}{mh}\left(1 -
      \frac{3E(m)}{2K(m)}\right)\right]\sqrt{gh},
\end{gather*}
where we added $h$ to $\eta$ to match the change in $\alpha$ from
$\sqrt{gh}$ to $-\frac{1}{2}\sqrt{gh}$ in (\ref{eq:dim:kdv}).
Here $K(m)$ and $E(m)$ are the complete elliptic integrals of the first
and second kind, respectively, and $\opn{cn}(z|m)$ is one of the
Jacobi elliptic functions \cite{dingemans,gradshteyn}.  In our
case $\lambda=2\pi$, $g=1$ and $h=0.05$.  For each $H$ in (\ref{eq:H:ABC}),
we solve the $\lambda$ equation for $m$ using Mathematica \cite{mma},
and then evaluate $\eta(x,0)$ on a uniform grid that is fine enough that
its Fourier coefficients decay below machine roundoff.  The values of
$m' = 1-m$ are
\begin{equation*}
  m'_A = 1.81924\times10^{-35}, \qquad
  m'_B = 1.98689\times10^{-14}, \qquad
  m'_C = 1.79643\times10^{-10}.
\end{equation*}
This approach requires extended precision arithmetic to compute $m$
and evaluate $\eta$, but the running time takes only a few seconds on
a typical laptop.  A periodized version of the simpler $\opn{sech}^2$
formula could be used for the first two waves, but decays too slowly
for wave $C$ to be spatially periodic to roundoff accuracy. Once these
cnoidal waves have been computed, we superpose their initial
conditions to form $\eta_0^{A+B}$ and $\eta_0^{B+C}$, just as in
\S\ref{sec:num}.  It is well-known that a superposition of $N$ cnoidal
waves retain this form when evolved via KdV, with $N$ amplitude and
$N$ phase parameters governed by an ODE describing pole dynamics in
the complex plane \cite{kruskal:pole,airault:kdv,deconinck:segur}. In
the $N=2$ case, the solutions are relative-periodic.

\begin{figure}[p]
\begin{center}
\includegraphics[width=.98\linewidth,trim=0 10 0 0]{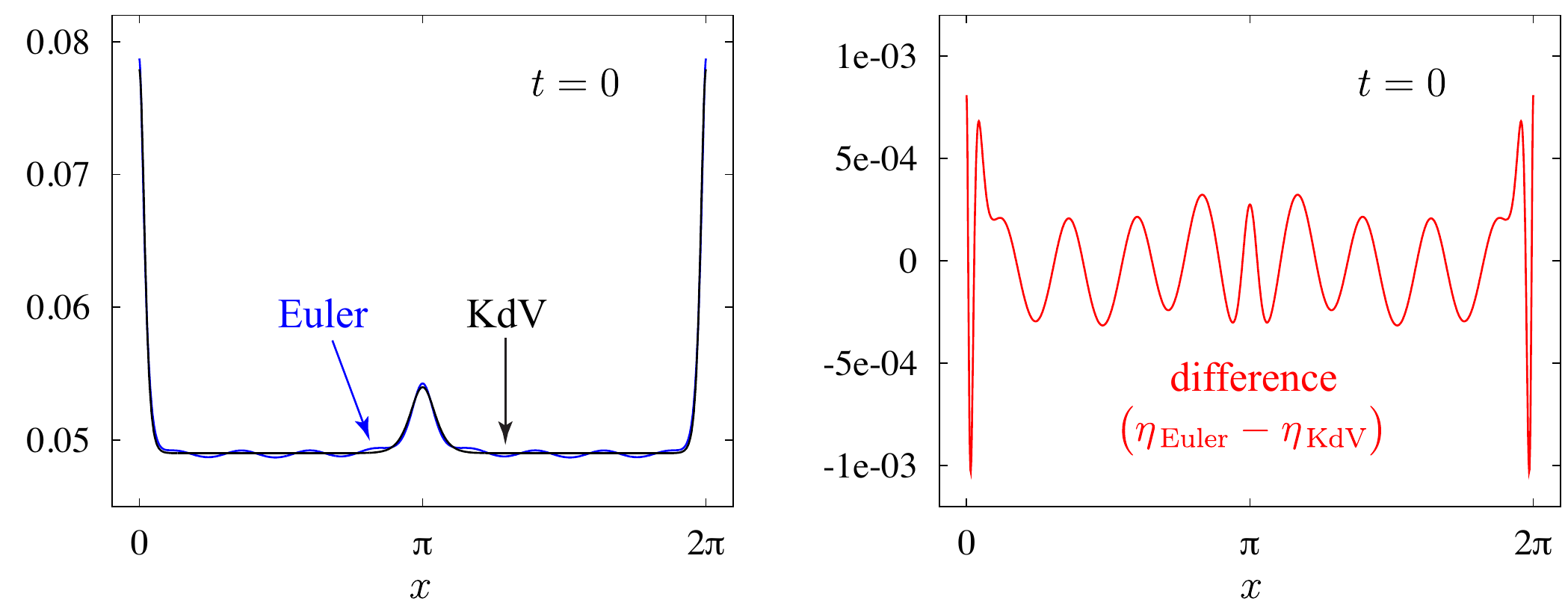}
\end{center}
\caption{\label{fig:kdv:cmp5} Comparison of time-periodic solution
  found in \S\ref{sec:num} to nearby relative-periodic two phase
  cnoidal solution of KdV. The periods are $T=138.387$ and $113.079$,
  respectively. }
\end{figure}

\begin{figure}[p]
\begin{center}
\includegraphics[width=.98\linewidth,trim=0 10 0 0]{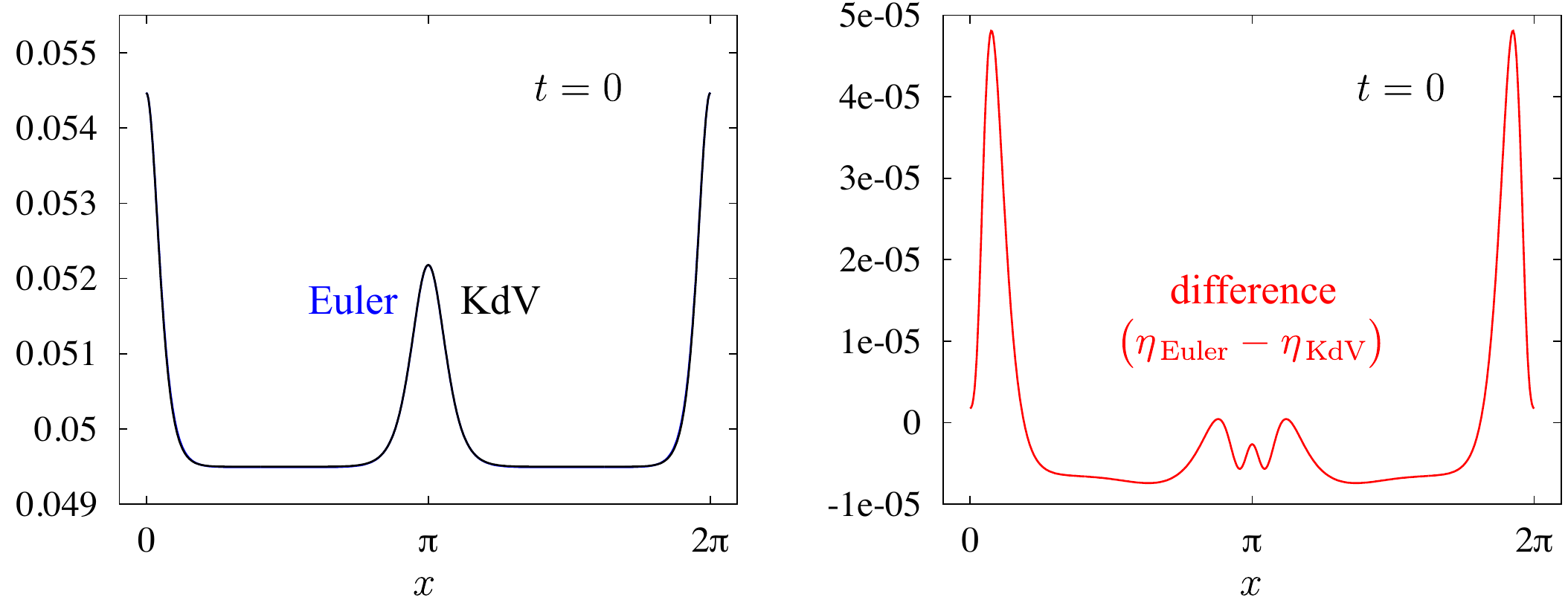}
\end{center}
\caption{\label{fig:kdv:cmp4} Comparison of relative-periodic solution
  found in \S\ref{sec:num} to nearby relative-periodic two phase
  cnoidal solution of KdV. The periods are $T=1108.11$ and $1068.73$,
  respectively. }
\end{figure}

Figures~\ref{fig:kdv:cmp5} and~\ref{fig:kdv:cmp4} compare the
time-periodic and relative-periodic solutions of the Euler equations,
computed in \S\ref{sec:num}, to these cnoidal solutions of KdV.  Since
the periods are different, only the initial conditions are compared.
In the larger-amplitude example, shown in Fig.~\ref{fig:kdv:cmp5}, the
Euler solution is not as flat in the wave trough as the cnoidal
solution due to an additional oscillatory component (the ``tuned''
radiation). From the difference plot in the right panel, we see that
the crest-to-trough amplitude of these higher frequency oscillations
is roughly $6\times10^{-4}$, or $2.1\%$ of the wave height $H_A$. The
Euler solution is time-periodic with period $T_\text{Euler}=138.387$
while the cnoidal solution is relative-periodic, returning to a
spatial phase shift of its initial condition at
$T_\text{KdV}=113.079$, which differs from $T_\text{Euler}$ by $18\%$.
In the smaller-amplitude example, shown in Fig.~\ref{fig:kdv:cmp4}, both
solutions have smooth, flat wave troughs, and it is difficult to
distinguish one from the other in the left panel.  The crest-to-trough
amplitude of the difference in the right panel is roughly
$5.5\times10^{-5}$, or $1.1\%$ of $H_B$. The relative change in period
is $(T_\text{Euler}-T_\text{KdV})/ T_\text{Euler} = 3.6\%$. While the
left panels of Figures~\ref{fig:kdv:cmp5} and~\ref{fig:kdv:cmp4} show
close agreement between relative-periodic solutions of the Euler and
KdV equations at $t=0$, it should be noted that the wave amplitudes of
the cnoidal solutions were chosen to minimize the discrepancy in these
figures.  The change in period by $18\%$ and $3.6\%$, respectively, is
perhaps a better measure of agreement.

\begin{figure}[p]
\begin{center}
\includegraphics[width=.98\linewidth,trim=0 10 0 0]{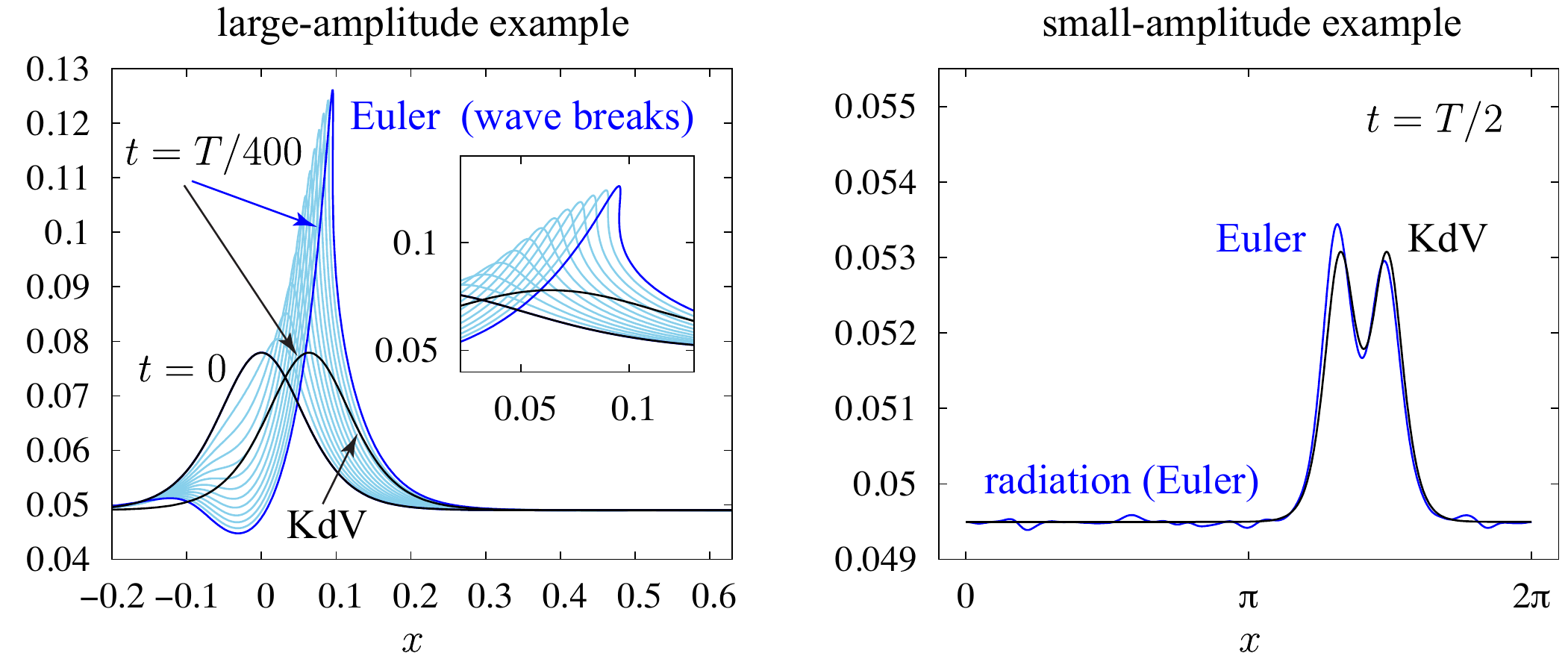}
\end{center}
\caption{\label{fig:kdv:cmp7} Comparison of KdV and Euler solutions,
  both initialized with a 2-phase cnoidal wave with peaks matching the
  heights of the Stokes waves labeled A and B (left) or B and C
  (right) in Fig.~\ref{fig:bif:stokes}. Here $T=138.387$ (left) and
  $T=1068.73$ (right). }
\end{figure}

A final comparison of the two equations is made in
Fig.~\ref{fig:kdv:cmp7}, where we evolve the Euler equations with the
KdV initial conditions.  This requires an initial condition for
$\varphi(x)=\phi(x,\eta(x))$, where we have suppressed $t$ in the
notation for this discussion since it is held fixed at 0.  Based on
the derivation presented above, we first solve $\phi_x(x,0) =
\sqrt{g/h}[\eta(x)-h]$ for $\phi$ on the bottom boundary.  We then use
the approximation
\begin{equation*}
  \varphi(x) \approx \phi(x,0) - \frac{\eta(x)^2}{2}\phi_{xx}(x,0) +
  \frac{\eta(x)^4}{24}\phi_{xxx}(x,0)
\end{equation*}
to evaluate $\phi$ on the free surface.  In the left panel of
Figure~\ref{fig:kdv:cmp7}, the larger wave grows and overturns before
$t=T/400$ when evolved under the Euler equations, instead of traveling
to the right when evolved via KdV.  To handle wave breaking, we
switched to an angle-arclength formulation of the free-surface Euler
equations \cite{hls94,vtxs1}.  In the small-amplitude example in the
right panel, the Euler solution develops visible radiation and falls
slightly behind the KdV solution, although the phases are closer at
$T/2$ than the result of evolving the Stokes waves under KdV in
Figure~\ref{fig:kdv:cmp2}.  We also tried evaluating
\begin{equation*}
  \phi(x,y)=\sqrt{g/h} \sum_{k=1}^\infty 2k^{-1}\hat\eta_k \sin(kx)\cosh(ky)
\end{equation*}
at $y=\eta(x)$ to obtain the initial condition for $\varphi(x)$, where
$\hat\eta_k$ are the Fourier modes of $\eta(x)$ at $t=0$, but the
results were worse for the large-amplitude example --- the wave
breaks more rapidly --- and were visually indistinguishable in the
small-amplitude example from the results plotted in
Fig.~\ref{fig:kdv:cmp7}.

In summary, the large-amplitude time-periodic solution of the Euler
equations found in \S\ref{sec:num} is well outside of the KdV regime
by any measure, and the small-amplitude relative-periodic solution is
closer, but not close enough to achieve quantitative agreement over
the entire time interval of interest.

\section{Conclusion}

We have demonstrated that the small amount of background radiation
produced when two Stokes waves interact in shallow water can often be
tuned to obtain time-periodic and relative-periodic solutions of the
free-surface Euler equations.  Just as for the Korteweg-de Vries
equation, the waves can fully merge when they collide or remain
well-separated.  However, the comparison is only qualitative as the
waves are too large to be well-approximated by KdV theory.

In future work, we will study the stability of these solutions using
Floquet theory. Preliminary results suggest that the first example
considered above is unstable to harmonic perturbations while the
second example is stable.  In the stable case, an interesting open
question is whether the Stokes waves used as a starting guess for the
minimization algorithm, which have the same energy as the
relative-periodic solution found, might remain close to it forever,
executing almost-periodic oscillations around it. Presumably $\theta$
would need to be varied slightly for this to be true, since $\theta$
is a free parameter that we selected by hand to obtain a small value
of $\tilde f$ for the initial guess.
Another open question is whether there are analogues for the Euler
equations of $N$-phase quasi-periodic solutions of the KdV equation
with $N\ge3$.  We are confident that the methods of this paper could
be used to construct degenerate cases of $N\ge3$ solitary water waves
colliding elastically in a time-periodic or relative-periodic fashion,
along the lines of what was done for the Benjamin-Ono equation in
\cite{benj2}.  Computing more general quasi-periodic dynamics of the
form $\eta(x,t)=H(\vec\kappa x + \vec\omega t + \vec\alpha)$,
$\varphi(x,t)=\Phi(\vec\kappa x + \vec\omega t + \vec\alpha)$ with
$H,\Phi\in C(\mathbb{T}^N)$ and $\vec\kappa$, $\vec\omega$,
$\vec\alpha\in\mathbb{R}^N$ seems possible in principle using a more
sophisticated shooting method to determine $H$, $\Phi$ and
$\vec\omega$.  Existence of such solutions for the Euler equations
would show that non-integrable equations can also support recurrent
elastic collisions even if they cannot be represented as $N$-phase
superpositions of elliptic functions.


\bibliographystyle{amsplain}
\providecommand{\bysame}{\leavevmode\hbox to3em{\hrulefill}\thinspace}
\providecommand{\MR}{\relax\ifhmode\unskip\space\fi MR }
\providecommand{\MRhref}[2]{%
  \href{http://www.ams.org/mathscinet-getitem?mr=#1}{#2}
}
\providecommand{\href}[2]{#2}

\end{document}